\documentclass[10pt,twocolumn,journal]{IEEEtran}
\usepackage{latexsym}
\usepackage{amsfonts}
\usepackage{amsbsy}
\usepackage{amsmath,amssymb}
\usepackage{times}
\usepackage{graphicx}
\usepackage{stkernel}
\usepackage{enumerate}
\usepackage[usenames]{color}
\usepackage[dvips]{pstcol}
\usepackage{epstopdf}
\usepackage{amsfonts,amssymb,amsmath,bm}
\usepackage{graphicx,subfigure}
\usepackage{epstopdf}
\usepackage{cite,color}
\usepackage{stfloats}
\usepackage{enumerate}
\usepackage{algorithm}
\usepackage{cite}
\usepackage[justification=raggedright,font={small}]{caption}
\newtheorem{theorem}{\textbf{Theorem}}
\newtheorem{lemma}{\textbf{Lemma}}

\newtheorem{assump}{\textbf{Assumption}}

\newcommand{\bI}{\mathbf{I}}

\newcommand{\bG}{\mathbf{G}}

\newcommand{\bA}{\mathbf{A}}

\newcommand{\bD}{\mathbf{D}}

\newcommand{\bg}{\bm{g}}

\newcommand{\ba}{\bm{a}}

\newcommand{\bx}{\bm{x}}

\newcommand{\by}{\bm{y}}
\newcommand{\bh}{\bm{h}}

\newcommand{\bb}{\bm{b}}
\newcommand{\bc}{\bm{c}}
\newcommand{\bd}{\bm{d}}
\newcommand{\bu}{\bm{u}}

\newcommand{\bn}{\bm{n}}
\newcommand{\Cdom}{\mathbb{C}}
\newcommand{\Rdom}{\mathbb{R}}

\newcommand{\cgauss}{\mathcal{CN}}
\newcommand{\opmin}{\mathop{\mathrm{minimize}}\limits}

\newcommand{\RNum}[1]{\uppercase\expandafter{\romannumeral #1\relax}}

\ifCLASSINFOpdf
\else
\fi
\begin{document}
\title{Simultaneously Transmitting and Reflecting (STAR) RIS Assisted Over-the-Air Computation Systems}
\author{Xiongfei Zhai, \IEEEmembership{Member,~IEEE,}
        Guojun Han, \IEEEmembership{Senior Member,~IEEE,}
        Yunlong Cai, \IEEEmembership{Senior Member,~IEEE,}\\
        Yuanwei Liu, \IEEEmembership{Senior Member,~IEEE,} and Lajos Hanzo, \IEEEmembership{Life Fellow,~IEEE}
        \thanks{
The work of X. Zhai was supported in part by the Science and Technology Program of Guangzhou under Grant 202102020869, and the Guangdong Basic and Applied Basic Research Foundation under Grants 2022A1515010153 and 2021A1515011645. The work of G. Han was supported in part by the Joint Funds of the National Natural Science Foundation of China and Guangdong under Grant U2001203, Key-Area R$\&$D Program of Guangdong Province under Grant 2021B1101270001, and Guangdong Provincial Key Laboratory of Photonics Information Technology under Grant 2020B121201011. The work of Y. Cai was supported in part by the National Natural Science Foundation of China under Grants 61971376 and U22A2004. This work of Y. Liu was supported in part by the CHIST-ERA grant under the project CHIST-ERA-20-SICT-005, by the Engineering and Physical Sciences Research Council (EPSRC) under Project EP/W035588/1, in part by the Royal Society under grant RGS$\backslash$R1$\backslash$221050 and grant IEC$\backslash$NSFC$\backslash$201112, and in part by the PHC Alliance Franco-British Joint Research Programme under Grant 822326028. L. Hanzo would like to acknowledge the financial support of the Engineering and Physical Sciences Research Council projects EP/W016605/1 and EP/P003990/1 (COALESCE) as well as of the European Research
Council's Advanced Fellow Grant QuantCom (Grant No. 789028). (Corresponding author: Yunlong Cai)

X. Zhai and G. Han are with the School of Information Engineering, Guangdong University of Technology, Guangzhou 510006, China (e-mail: zhaixiongfei@gdut.edu.cn; gjhan@gdut.edu.cn).

Y. Cai is with the College of Information Science and Electronic Engineering, Zhejiang University, Hangzhou 310027, China (e-mail: ylcai@zju.edu.cn).

Y. Liu is with the School of Electronic Engineering and Computer Science, Queen Mary University of London, London E1 4NS, UK, (email: yuanwei.liu@qmul.ac.uk).

L. Hanzo is with the Department of Electronics and Computer Science, University of Southampton, Southampton, UK (Email: lh@ecs.soton.ac.uk).
}
}
\maketitle
\begin{abstract}
The performance of over-the-air computation (AirComp) systems degrades
due to the hostile channel conditions of wireless devices (WDs), which
can be significantly improved by the employment of reconfigurable
intelligent surfaces (RISs). However, the conventional RISs require
that the WDs have to be located in the half-plane of the reflection
space, which restricts their potential benefits. To address this
issue, the novel family of simultaneously transmitting and reflecting
reconfigurable intelligent surfaces (STAR-RIS) is considered in
AirComp systems to improve the computation accuracy across a wide
coverage area. To minimize the computation mean-squared-error (MSE) in
STAR-RIS assisted AirComp systems, we propose a joint beamforming
design for optimizing both the transmit power at the WDs, as well as
the passive reflect and transmit beamforming matrices at the STAR-RIS,
and the receive beamforming vector at the fusion center
(FC). Specifically, in the updates of the passive reflect and transmit
beamforming matrices, closed-form solutions are derived by introducing
an auxiliary variable and exploiting the coupled binary phase-shift
conditions. Moreover, by assuming that the number of antennas at the
FC and that of elements at the STAR-RIS/RIS are sufficiently high, we
theoretically prove that the STAR-RIS assisted AirComp systems provide
higher computation accuracy than the conventional RIS assisted
systems. Our numerical results show that the proposed beamforming
design outperforms the benchmark schemes relying on random phase-shift
constraints and the deployment of conventional RIS. Moreover, its
performance is close to the lower bound achieved by the beamforming
design based on the STAR-RIS dispensing with coupled phase-shift
constraints.
\end{abstract}
\IEEEpeerreviewmaketitle
\section{Introduction}
The ultra-fast aggregation of the deluge of data generated by
distributed wireless devices (WDs) will become an urgent demand for
the Internet-of-things
(IoT)~\cite{Agiwal2016,Lin2017,Chen2020}. However, due to the fact
that the fusion center (FC) of conventional communication systems has
to recover the individual data stream of each WD and then perform data
aggregation, there are three dominant challenges: 1) the inter-WD
interference will degrade the performance of data stream recovery; 2)
the successive data stream recovery will lead to extremely high
latency; 3) the spectral- and energy-efficiency will be severely
affected by the increasing number of WDs. Since the FC typically has
to compute a specific function (e.g., the arithmetic mean, the
weighted sum, or the geometric mean) in data aggregation, the
compelling technique of over-the-air computation (AirComp) has been
proposed as a potential
remedy~\cite{Cao2019,Liu2020,Xiao2008,Wang2011,ZhuAir1,Zang2020,Wen2019,Abari2015}. However,
as the the number of WDs increases rapidly, the computation accuracy
of AirComp tends to erode, since it is challenging to design a single
receiver for the FC to aggregate the data from an escalating number of
WDs. To address this issue, advanced massive multiple-input
multiple-output (MIMO) and reconfigurable intelligent surface (RIS)
aide techniques may be harnessed in AirComp. Specifically, the RIS is
more attractive due to its flexible deployment and reduced power
consumption. Although the traditional RISs beneficially improve the
computational performance of AirComp, they can only provide half-plane
coverage, while the WDs are spread across the entire 360-degree
angular range. To extend the coverage, the recent simultaneously
transmitting and reflecting RISs (STAR-RISs) have excellent potential
in AirComp due to the fact that they can reflect and transmit the
incident signals simultaneously. Therefore, full-plane coverage is
possible in the STAR-RIS assisted AirComp systems.


\subsection{Prior Works}
In the literature, the computation mean-square-error (MSE) metric is
the most popular one in AirComp systems
\cite{Cao2019,Liu2020,Xiao2008,Wang2011,ZhuAir1,Zang2020,Wen2019,Abari2015}. By
assuming fading channels and single-antenna FC/WDs, the power
allocation algorithms of AirComp systems based on popular convex
optimization methods have been investigated in
\cite{Cao2019,Liu2020,Xiao2008}. Instead of minimizing the computation
MSE, a new computation metric, namely the outage probability, is
defined as the probability that the computation MSE exceeds a given
threshold and is minimized in~\cite{Wang2011}. However, the
aforementioned contributions focused their attention on single-input
single-output (SISO) AirComp systems, which cannot support
multi-functional computation. To deal with this issue, the authors
of~\cite{ZhuAir1} exploited the MIMO techniques in their AirComp
systems. Specifically, a zero-forcing (ZF) based beamforming design
was proposed for MIMO-aided AirComp systems in~\cite{ZhuAir1} for
facilitating multi-functional computations and for minimizing the
computation MSE. To guarantee low computation MSE, accurate time
synchronization is important for AirComp systems. To achieve time
synchronization both among the WDs as well as between the FC and WDs,
a synchronization scheme, which is referred to as AirShare, has been
developed in~\cite{Abari2015}. In this scheme, the FC broadcasts a
synchronization block to all WDs for time synchronization.

Nevertheless, the computation accuracy of AirComp will be severely
degraded by the rapidly increasing number of WDs due to the ones
having poor channel conditions. To overcome this challenge, one of the
effective techniques is to provide higher spatial degrees. Thus the
massive MIMO techniques were exploited for AirComp
in~\cite{Wen2019,Zhai2020,Jeon2020}. In particular, in our previous
work~\cite{Zhai2020} we conceived a hybrid analog-digital (AD)
beamformer for massive MIMO AirComp systems and proposed a
successive-convex-approximation (SCA) based beamforming design to
achieve low computation MSE. Moreover, the computation MSE has been
proven to be inversely proportional to the number of receive antennas
at the FC. Although the computation accuracy can be significantly
improved by massive MIMO schemes, this is generally at the cost of
huge energy consumption and complexity. Thus finding a more efficient
scheme for enhancing the computation performance of AirComp is of
pivotal importance.

Again, the passive RIS concept has attracted substantial attention
worldwide due to its compelling capability of increasing the spatial
degrees of freedom, whilst relying on passive reflecting elements
\cite{Wu2019,Wu2020,Huang2019,Renzo2019,Basar2019}. By adjusting the
phase shift and even potentially the amplitude of the incident signal
with the aid of passive reflecting elements, the propagation environment
is beneficially influenced. As compared to massive MIMO, the
fabrication cost and energy consumption of RIS is modest, since
no radio frequency (RF) chains are used at the
RIS. Given the above advantages, RISs have been exploited in
diverse applications, such as RIS-aided wireless power
transfer~\cite{WuWCL2020,WuJ2020}, RIS assisted physical layer
security~\cite{Cui2019,Xu2019,Guan2020,Jiang2020} and RIS-aided orthogonal
frequency division multiplexing (OFDM)~\cite{Yang2020, Zheng2020}. Specifically, the authors in \cite{WuJ2020} investigated the RIS-assisted simultaneous wireless information and power transfer (SWIPT) systems with the quality-of-service (QoS) constraints at the information and energy users. A penalty-based algorithm was proposed to address the non-convexity of the corresponding optimization problem. To achieve the secure transmission, the RIS-assisted MIMO systems was studied in \cite{Jiang2020} and an alternating optimization algorithm with SCA was developed to jointly optimize the beamforming design at the transceiver and RIS. To apply the RIS in the frequency-selected channels, the combination of RIS and OFDM was exploited in \cite{Yang2020}. To estimate the channel state information (CSI) with reduced overhead, a RIS-element-grouping method was proposed. Then the system rate maximization problem was formulated and solved by alternately optimizing the power allocation and passive array coefficients.

RISs have also been exploited in AirComp
systems~\cite{Jiang2019,Yu2020,Wang2021,ZhaiIoT2021}. In particular, a
joint active and passive beamforming design based on
difference-of-convex (DC) programming and matrix lifting has been
proposed in~\cite{Jiang2019} for minimizing the computation
MSE. Furthermore, the authors of~\cite{Wang2021} combined AirComp and
energy-harvesting, where the RIS-assisted beamforming design problem
is separated into two phases: the energy beamforming phase and the
AirComp phase. To reduce the heavy signaling overhead due to the
estimation of CSI, a two-stage stochastic
beamforming algorithm has been proposed in~\cite{ZhaiIoT2021}, where
the passive beamforming matrix employed at the RIS is optimized based
on the channel statistics, while the transmit power at the WDs and the
receive beamforming vector at the FC are updated with the aid of the
effective near-real-time CSI. Although the RISs can significantly
improve the computation performance of AirComp, the aforementioned
treatises only considered conventional RISs, which can only reflect
the incident signals to a 180-degree half-plane. In other words, the
FC and WDs have to be located on the same side of the RIS. Hence this
geographical constraint restricts the efficiency, appeal and
flexibility of RISs, especially when the FC and WDs are located on
both sides of a RIS on a 360-degree disc.

To address this issue, a novel technique, which is referred to as
STAR-RISs, has been developed in~\cite{Xu2021CL} and
\cite{LiuWC2021}. In particular, the incident signals impinging on the
elements of STAR-RIS from either side will be divided into two parts,
where one of them is reflected to the same side of the incident
signals into the so-called {\em reflection space}, while the other is
transmitted to the opposite side, namely to the {\em transmission
  space}. This can be achieved by adjusting the electric and magnetic
currents of the elements at the STAR-RIS. Thus the incident signals
can be transmitted and reflected with the aid of two types of
coefficients, which are referred to as the transmission and reflection
coefficients. Some prototypes of STAR-RISs have been proposed in the
literature based on metasurfaces~\cite{Zhu2014,Docomo2020}. The
authors of~\cite{Zhu2014} considered the passive elements with a
parallel resonant LC tank and small metallic loops, which can provide
the required electric and magnetic surface reactance. Due to the
aforementioned advantages, some authors have also considered the
exploitation of STAR-RIS in wireless communication
systems~\cite{LiuArX2021Oct,Mu2021,Niu2022}. In particular, three
different operating protocols have been developed for STAR-RISs
in~\cite{Mu2021}, namely energy splitting, mode switching, and time
switching. As a further advance, the authors of~\cite{Mu2021} proposed
the joint active and passive beamforming concept for minimizing the
power consumption of STAR-RIS aided unicast and multicast systems. By
considering the coupled phase-shift constraints between the transmit
and reflect matrices at the STAR-RIS, the authors
of~\cite{LiuArX2021Oct} investigated the power consumption
minimization problems under specific user-rate constraints for both
non-orthogonal multiple access (NOMA) and orthogonal multiple access
(OMA) systems by relying on an element-wise alternating
optimization algorithm.

\subsection{Motivations and Contributions}
However, to the best of our knowledge, the application of STAR-RISs in
AirComp systems and the corresponding beamforming design have not been
investigated in the open literature. Hence closing this knowledge-gap
is the main motivation of this work. Furthermore, as compared to
traditional RISs, which can only provide half-plane coverage for the
AirComp systems, STAR-RISs is capable of covering the full
plane. Furthermore, the limited coverage of the traditional RISs
introduces increased design/coordination complexity, especially for a dense
network, while STAR-RISs can be regarded as being transparent to the
communication environment. Our main contributions of this paper are
summarized as follows:

\begin{itemize}
\item\; We propose a STAR-RIS assisted AirComp system, where a
  multi-antenna FC aims for aggregating the signals from a number of
  single-antenna WDs with the aid of a STAR-RIS. In contrast to the
  traditional RIS-assisted AirComp systems, the WDs located in both
  the reflection and transmission spaces may be supported. Thus the
  transmit power at the WDs, the receive beamforming vector at the FC,
  and the passive reflect/transmit beamforming matrices at the
  STAR-RIS have to be jointly optimized for minimizing the computation
  MSE. No research on such systems has been reported in the open
  literature as yet. Besides, the formulated problem is challenging due to the fact that the passive reflect and transmit beamforming matrices at the STAR-RIS are coupled in the phases and are both
subjected to the non-convex unit modulus constraints.
\item\; An efficient optimization algorithm is proposed for updating the
  variables in an alternating manner. First, the receive beamforming
  vector at the FC and the transmit power at the WDs are optimized by
  exploiting the first-order optimality condition and the Lagrange
  duality method, respectively. Then we reformulate the original
  problem by introducing an auxiliary binary vector and exploiting the
  binary phase-coupled constraints. The closed-form solutions of the
  transmit/reflect beamforming matrices at the STAR-RIS are derived
  based on the constant modulus constraints and the binary nature. The
  complexity and convergence of the proposed algorithm are also
  analyzed.
\item\; Given a sufficiently high numbers of antennas at the FC and
  elements at the STAR-RIS, we theoretically derive approximate
  expression for the computation MSEs of both the STAR-RIS assisted and
  conventional RIS assisted systems. Then it is proven that the
  STAR-RIS assisted AirComp systems can achieves lower computation MSE
  than their conventional RIS assisted AirComp counterparts, which further validate the effectiveness of STAR-RIS.
\item\; Our numerical results verify the convergence of the proposed
  algorithm. It is also shown that our proposed algorithm outperforms
  other benchmark schemes based on the random phase-shift constraints
  and conventional RIS. Thus our algorithm achieves both reduced
  computation MSE and full-plane coverage compared to its conventional
  RIS-aided counterpart. Moreover, the computation MSE achieved by the
  proposed algorithm is close to the lower bound attained by the
  scheme based on STAR-RIS without coupled phase-shift
  constraints.
\end{itemize}

\subsection{Organization and Notations}
The remainder of this paper is organized as follows. Section~\RNum{2}
introduces the system model and the problem formulated. The proposed
joint beamforming design and the analysis of its convergence and
complexity are presented in Section~\RNum{3}. The comparison between
the STAR-RIS assisted AirComp systems and their conventional RIS-aided
counterparts are shown in Section~\RNum{4}. Finally, Section~\RNum{5}
presents our simulation results, followed by our conclusions in
Section~\RNum{6}.

{\it Notations}: Throughout this paper, we adopt bold upper-case
letters for matrices and bold lower-case letters for
vectors. $\bA(i,j)$ denotes the entry on the $i^{th}$ row and the
$j^{th}$ column for a matrix $\bA$, while $\bA^T$, $\bA^H$, $\bA^{*}$,
and $\bA^{-1}$ denote its transpose, Hermitian transpose, conjugate,
and inverse, respectively. We denote $\bI$ as the identity matrix
whose dimension will be clarified from the context, and denote
$\Cdom^{m\times n}$ ($\Rdom^{m\times n}$) as the $m$-by-$n$
dimensional complex (real) space. The notations $\mathbb{E}(\cdot)$,
$|\cdot|$, $\Re(\cdot)$, and $\text{sign}(\cdot)$ represent the
expectation, absolute value, Real part, and sign of an input variable,
respectively. We denote $\circ$ as the Hadamard product between two
matrices. $\cgauss(\bm{\Upsilon},\bm{\Phi})$ denotes the circularly
symmetric complex Gaussian (CSCG) distribution with mean
$\bm{\Upsilon}$ and covariance matrix $\bm{\Phi}$.

\section{System Description}
\begin{figure}[htbp]
\centering
\includegraphics[width=2.5in]{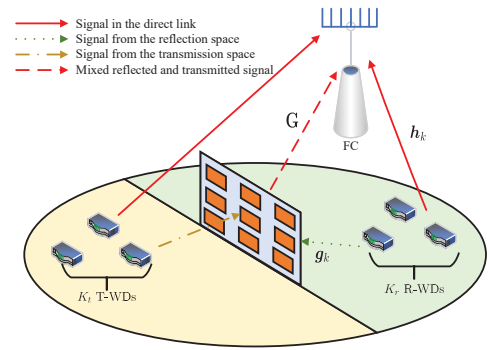}
\captionsetup{justification=raggedright}
\caption{A STAR-RIS assisted AirComp system.}
\label{fig:fig1}
\end{figure}
In this section, we first introduce the investigated STAR-RIS assisted AirComp system. Then the optimization problem is formulated to minimize the computation MSE.

\subsection{System Model}
In this paper, we study a STAR-RIS assisted AirComp system as shown in Fig. \ref{fig:fig1}, which consists of a multi-antenna FC, a STAR-RIS, and $K$ single-antenna WDs. We assume that the FC is equipped with an uniform linear array (ULA) and the number of antennas is $N$. In this system, $K$ WDs first collect one heterogeneous time-varying parameter from the environment (e.g., temperature, noise, humidity), and then transmit the collected parameter to the FC via wireless channels for functional computation. We employ a STAR-RIS with $M$ passive reflecting/transmitting elements to improve the computation MSE performance. In the conventional RIS assisted AirComp system, all WDs should be at the same side of the RIS due to the fact that the RIS can only perform signal reflection. However, the STAR-RIS is also able to transmit the incident signal, allowing that the WDs can be located at the different sides. Hence, the deployment of the STAR-RIS is much more flexible than the conventional one. Specifically, the space in Fig. \ref{fig:fig1} is separated into two parts by the STAR-RIS, i.e., the reflection space and the transmission space. Similarly, the WDs are also divided into two groups, called the R-WD and T-WD groups, which are located in the reflection and transmission space, respectively. We assume that the numbers of R-WDs and T-WDs are $K_r$ and $K_t$ with $K_r+K_t=K$. Therefore, the STAR-RIS can reflect the signal from the reflection space (the R-WDs) as well as transmit the signal from the transmission space (the T-WDs) after the processing of passive beamforming. For convenience, we respectively denote $\mathcal{K}\triangleq \{1,2,\ldots,K\}$, $\mathcal{K}_r\triangleq \{1,2,\ldots,K_r\}$, $\mathcal{K}_t\triangleq \{K_r+1,\ldots, K\}$, and $\mathcal{M}\triangleq \{1,2,\ldots,M\}$ as the sets of all WDs, R-WDs, T-WDs, and the passive reflect/transmit elements at the STAR-RIS. By applying the AirShare technique in \cite{Abari2015}, we assume that the synchronization of different WDs is established.

At a particular time slot, let $s_k\in\Cdom$ denote the collected parameter by WD $k$. We would like to note that the parameter $s_k$ is assumed to be normalized and independent of those from different WDs, i.e., $\mathbb{E}(s_ks_k^{*})=1$ and $\mathbb{E}(s_ks_j^{*})=0$, $k,j\in\mathcal{K}$. For swift data aggregation, the FC aims at computing the sum of collected parameters from all the WDs (i.e., $s=\sum_{k\in\mathcal{K}}s_k$). Besides, the design investigated in this paper can be readily extended to other nomographic functions \cite{Goldenbaum2015}. By denoting $v_k$ as the transmit coefficient at WD $k$, then the corresponding transmitted data is given by
\begin{align}
x_k = v_ks_k.
\end{align}
Let $P$ denote the transmit power budget and the transmit power constraints at the WDs is expressed as $|v_k|^2\leq P,k\in\mathcal{K}$. By considering the deployment of the STAR-RIS, the incident signals at the STAR-RIS from two groups of WDs, i.e., the R-WDs and T-WDs, are written as
\begin{align}
&\by_r = \sum_{k\in\mathcal{K}_r}\bg_kv_ks_k,\\
&\by_t = \sum_{k\in\mathcal{K}_t}\bg_kv_ks_k,
\end{align}
respectively, where $\bg_k\in\Cdom^{M\times 1}$ denotes the channel vector from WD $k$ to the STAR-RIS. In this paper, we assume that the STAR-RIS operates based on the energy splitting protocol \cite{Mu2021}~\footnote{In this work, for ease of analysis, the CSI related to the WDs, the STAR-RIS, and the FC is assumed to be known. This can be achieved by exploiting the uplink/downlink training methods in \cite{Zheng2020,Rao2014,Soltanalian2017,Goldenbaum2014}. A robust beamforming design can be developed by following the previous robust design techniques \cite{Robust0,Robust1,Robust2}, which will be addressed in our future research.} In this protocol, a single passive beamforming matrix at the conventional RIS is divided into two parts at the STAR-RIS. One is designed for signal reflection, namely the passive reflect beamforming matrix, while the other is used for signal transmission, called the passive transmit beamforming matrix. Then the passive reflect and transmit beamforming matrices are denoted by $\bm{\Theta}_r\triangleq\sqrt{\beta_r}\text{diag}\left(\bm{\theta}_r^{*}\right)\in\Cdom^{M\times M}$ and $\bm{\Theta}_t\triangleq\sqrt{\beta_t}\text{diag}\left(\bm{\theta}_t^{*}\right)\in\Cdom^{M\times M}$, respectively, where $\beta_r,\beta_t\in[0,1]$, $\bm{\theta}_r\triangleq[e^{j\theta_1^r},e^{j\theta_2^r},\ldots,e^{j\theta_M^r}]^T\in\Cdom^{M\times 1}$ and $\bm{\theta}_t\triangleq[e^{j\theta_1^t},e^{j\theta_2^t},\ldots,e^{j\theta_M^t}]^T\in\Cdom^{M\times 1}$ represent the diagonal reflect and transmit beamforming vectors, $\theta_m^r$ and $\theta_m^t$ denote the reflection and transmission phase-shift coefficients, respectively, with $\theta_m^r,\theta_m^t\in[0,2\pi]$, $\forall m\in\mathcal{M}$.\footnote{The continuous phase adjustment is assumed at the RIS's elements for ease of analysis. However, we would like to note that our proposed beamforming design can be also applied to the cases with the discrete phase adjustment. This can be achieved by projecting the continuous phase to the nearest element of the discrete-phase set. } When $\beta_t=0$, the considered STAR-RIS reduces to a conventional RIS. Hence, the latter can be viewed as a special case. Since the correlation between $\beta_r$ and $\beta_t$ makes the system model and the corresponding problem much more complicated, to avoid high complexity and show more insights, we consider a simplified equal-energy splitting protocol in our initial attempt, i.e., $\beta_r=\beta_t=0.5$. Furthermore, for ease of presentation, $\beta_r$ and $\beta_t$ are both scaled up to be $1$, which means the STAR-RIS can perform full-power signal reflection and transmission simultaneously. To perform the signal reflection and transmission, the passive elements at the STAR-RIS need to support both electric and magnetic currents \cite{LiuWC2021,Zhu2014}. Hence, we consider a coupled phase-shift model for STAR-RIS, which leads to the following condition for the phase-shift coefficients of the passive elements \cite{LiuArX2021Oct}:
\begin{align}
\cos(\theta_m^t-\theta_m^r)=0, \forall m\in\mathcal{M}.
\end{align}
This condition shows that the reflection and transmission phase-shift coefficients are coupled one by one and satisfy the constraint $|\theta_m^t-\theta_m^r|=\frac{\pi}{2},\frac{3\pi}{2}, \forall m\in\mathcal{M}$. By jointly optimizing the passive reflect and transmit beamforming matrices at the STAR-RIS, we can obtain a high degree of flexibility for reducing the computation MSE in STAR-RIS assisted AirComp systems. After the processing of the passive reflect and transmit beamforming matrices, the mixed reflected and transmitted signal by the STAR-RIS is given by
\begin{align}
\bx_{\text{rt}} = \bm{\Theta}_r\sum_{k\in\mathcal{K}_r}\bg_kv_ks_k+\bm{\Theta}_t\sum_{k\in\mathcal{K}_t}\bg_kv_ks_k.
\end{align}

Hence, the received signal at the FC can be expressed as
\begin{align}
\by =& \underbrace{\sum_{k\in\mathcal{K}_r}\left(\bh_k+\bG\bm{\Theta}_r\bg_k\right)v_ks_k}_{\text{Signal from the reflection space}}+\underbrace{\sum_{k\in\mathcal{K}_t}\left(\bh_k+\bG\bm{\Theta}_t\bg_k\right)v_ks_k}_{\text{Signal from the transmission space}}\nonumber\\
&+\bn,
\end{align}
where $\bh_k\in\Cdom^{N\times 1}$, $\bG\in\Cdom^{N\times M}$, and $\bn\in\Cdom^{N\times 1}$ denote the channel vector from WD $k$ to the FC, the channel matrix from the STAR-RIS to the FC, and the additive white Gaussian noise (AWGN) vector with $\mathbb{E}(\bn)=\bm{0}$ and $\mathbb{E}(\|\bn\|^2)=\sigma^2\bI$, respectively. As we can see, the received signal at the FC can be separated into three parts, i.e., the signal from the reflection space, the signal from the transmission space, and the noise vector. Then the FC aims to compute the sum of the transmitted parameters from all the WDs (in the reflection and transmission spaces). Let $\bu\in\Cdom^{N\times 1}$ denote the receive beamforming vector at the FC, thus the obtained result after computing is expressed as
\begin{align}
\hat{s}=\bu^H\by.
\end{align}

\subsection{Problem Formulation}
With the above derivation, the computation performance can be measured by the MSE between $s$ and $\hat{s}$, namely the computation MSE, which is given by
\begin{align}
\zeta &= \mathbb{E}(|\hat{s}-s|^2)\nonumber\\
&=\sum_{k\in\mathcal{K}_r}|\bu^H(\bh_k+\bG\bm{\Theta}_r\bg_k)v_k-1|^2\nonumber\\
& + \sum_{k\in\mathcal{K}_t}|\bu^H(\bh_k+\bG\bm{\Theta}_t\bg_k)v_k-1|^2+\sigma^2\bu^H\bu\nonumber\\
&=\sum_{k\in\mathcal{K}_r}|\bu^H\bh_kv_k+\bm{\theta}_r^H\text{diag}(\bu^H\bG)\bg_kv_k-1|^2\nonumber\\
&+\sum_{k\in\mathcal{K}_t}|\bu^H\bh_kv_k+\bm{\theta}_t^H\text{diag}(\bu^H\bG)\bg_kv_k-1|^2+\sigma^{2}\bu^H\bu.
\end{align}
In this work, we focus on minimizing the computation MSE by optimizing the transmit coefficients at the WDs, the receive beamforming vectors at the FC, and the diagonal reflect/transmit beamforming vectors at the STAR-RIS, i.e., solving the following problem:
\begin{equation}\label{Oripro}
\begin{split}
\mathcal{P}1\!:\;&\opmin_{\{v_k\},\bu,\bm{\theta}_r,\bm{\theta}_t} ~ \zeta\\
&\text{subject to} ~ |v_k|^2\leq P,\forall k\in\mathcal{K},\\
&~~~~~~~~~~~~\bm{\theta}_r(m)=e^{j\theta_m^r},\forall m\in\mathcal{M},\\
&~~~~~~~~~~~~\bm{\theta}_t(m)=e^{j\theta_m^t},\forall m\in\mathcal{M},\\
&~~~~~~~~~~~~|\theta_m^t-\theta_m^r|=\frac{\pi}{2}~\text{or}~\frac{3\pi}{2},\forall m\in\mathcal{M},\\
&~~~~~~~~~~~~\theta_m^r,\theta_m^t\in[0,2\pi],\forall m\in\mathcal{M}.
\end{split}
\end{equation}
One can see that, there are two kinds of constraints in problem $\mathcal{P}1$, i.e., the transmit power constrains at the WDs and the structure constraints at the STAR-RIS. Specifically, the second and third constraints are constant modulus constraints due to the structure of STAR-RIS, while the coupled phase-shift condition leads to the last constraint. Multiple variables are coupled in the non-convex and non-linear constraints, which makes problem $\mathcal{P}1$ more challenging. To the best of our knowledge, there lacks efficient algorithms to solve this non-convex beamforming design problem.

\section{The Proposed Alternating Optimization Beamforming Design}
In this section, we propose an alternating-optimization (AO) algorithm based on the binary phase-shift constraints (BPCs), namely AO-BPC, to solve problem $\mathcal{P}1$. In particular, we first optimize the receive beamforming vector at the FC by fixing the transmit coefficients at the WDs and the diagonal reflect/transmit beamforming vectors at the STAR-RIS based on the first-order optimality condition. Then the transmit coefficients are updated given the receive beamforming vector and the diagonal reflect/transmit beamforming vectors by exploiting the Lagrange duality method. Finally, the diagonal reflect and transmit beamforming vectors are decoupled by introducing an auxiliary binary vector and exploiting the binary phase-shift constraints. Then the closed-form solutions are derived based on the constant modulus constraints and the binary nature. The computational complexity and convergence of the proposed algorithm are both analyzed.

\subsection{Optimization of Receive Beamforming Vector $\bu$}
By fixing $\{v_k\}$, $\bm{\theta}_r$, and $\bm{\theta}_t$, the subproblem with respect to the receive beamforming vector at the FC is given by
\begin{equation}\label{upro}
\mathcal{P}2\!:\;\opmin_{\bu} ~ \zeta.
\end{equation}
Obviously, problem $\mathcal{P}2$ is unconstrained and convex, which can be solved by checking the first-order optimality condition. Therefore, the optimal solution for $\bu$ is expressed as
\begin{align}
\bu =& \left(\sum_{k\in\mathcal{K}_r}|v_k|^2\bh_{k,r}\bh_{k,r}^H+\sum_{k\in\mathcal{K}_t}|v_k|^2\bh_{k,t}\bh_{k,t}^H+\sigma^2\bI\right)^{-1}\nonumber\\
&\times\left(\sum_{k\in\mathcal{K}_r}\bh_{k,r}v_k+\sum_{k\in\mathcal{K}_t}\bh_{k,t}v_k\right),\label{optu}
\end{align}
where $\bh_{k,r}=\bh_k+\bG\bm{\Theta}_r\bg_k,k\in\mathcal{K}_r$ and $\bh_{k,t}=\bh_k+\bG\bm{\Theta}_t\bg_k,k\in\mathcal{K}_t$ denote the equivalent channel vectors from the R-WDs and T-WDs to the FC, respectively. Compared to the sum-MMSE receive beamforming in conventional RIS assisted AirComp systems, which is in the form of $\left(\sum_{k\in\mathcal{K}}|v_k|^2\bh_{k}\bh_{k}^H+\sigma^2\bI\right)^{-1}\sum_{k\in\mathcal{K}}\bh_{k}v_k$, the terms inside and outside the matrix inversion are quite different due to the fact that the R-WDs and T-WDs transmit signal to the FC via different paths.

\subsection{Optimization of the Transmit Coefficients $\{v_k\}$}
With given $\bu$, $\bm{\theta}_r$, and $\bm{\theta}_t$, we can formulate the subproblem with respect to $\{v_k\}$ as follows:
\begin{equation}\label{vkpro}
\begin{split}
\mathcal{P}3\!:\;&\opmin_{\{v_k\}} ~ \zeta\\
&\text{subject to} ~ |v_k|^2\leq P,\forall k\in\mathcal{K}.
\end{split}
\end{equation}
It can be observed that the transmit coefficients at the R-WDs and those at the T-WDs are separated in the objective function and constraint. Furthermore, the transmit coefficients among the R-WDs/T-WDs are also separated. Hence, we can solve problem $\mathcal{P}3$ by optimizing $v_k$ in the order of $k$. Then the problem with respect to the transmit coefficient at R-WD $k$ can be formulated as
\begin{equation}\label{vkpro0}
\begin{split}
\mathcal{P}4\!:\;&\opmin_{v_k} ~ |\bu^H\bh_{k,r}v_k-1|^2\\
&\text{subject to} ~ |v_k|^2\leq P.
\end{split}
\end{equation}
Similarly, the transmit coefficient optimization problem for T-WD $k$ is given by
\begin{equation}\label{vkpro1}
\begin{split}
\mathcal{P}5\!:\;&\opmin_{v_k} ~ |\bu^H\bh_{k,t}v_k-1|^2\\
&\text{subject to} ~ |v_k|^2\leq P.
\end{split}
\end{equation}
Problem $\mathcal{P}4$ and $\mathcal{P}5$ are both convex problems with one quadratic constraint. Hence, the Lagrange duality method can be exploited to solve these two problems efficiently, leading to the following lemma.
\lemma The optimal solutions for problem $\mathcal{P}4$ and $\mathcal{P}5$ are given by
\begin{align}
&v_k^{\star} = \frac{\bh_{k,r}^H\bu}{\bu^H\bh_{k,r}\bh_{k,r}^H\bu+\mu_{k,r}^{\star}},\forall k\in\mathcal{K}_r,\\
&v_k^{\star} = \frac{\bh_{k,t}^H\bu}{\bu^H\bh_{k,r}\bh_{k,t}^H\bu+\mu_{k,t}^{\star}},\forall k\in\mathcal{K}_t,
\end{align}
respectively, where $\mu_{k,r}^{\star}$ and $\mu_{k,t}^{\star}$ denote the optimal Lagrange multipliers associated with the transmit power constraints in problem $\mathcal{P}4$ and $\mathcal{P}5$. If $\bu^H\bh_{k,r}\bh_{k,r}^H\bu$ and $\bu^H\bh_{k,t}\bh_{k,t}^H\bu$ are non-zero and
\begin{align}
&\frac{1}{\bu^H\bh_{k,r}\bh_{k,r}^H\bu}<P, \forall k\in\mathcal{K}_r,\\
&\frac{1}{\bu^H\bh_{k,t}\bh_{k,t}^H\bu}<P,\forall k\in\mathcal{K}_t,
\end{align}
we choose $\mu_{k,r}^{\star}=0$ and $\mu_{k,t}^{\star}=0$; otherwise, we choose $\mu_{k,r}^{\star}$ and $\mu_{k,t}^{\star}$ such that the following conditions are satisfied
\begin{align}
&\left|\frac{\bu^H\bh_{k,r}}{\bu^H\bh_{k,r}\bh_{k,r}^H\bu+\mu_{k,r}^{\star}}\right|^2=P,\forall k\in\mathcal{K}_r,\\
&\left|\frac{\bu^H\bh_{k,t}}{\bu^H\bh_{k,t}\bh_{k,t}^H\bu+\mu_{k,t}^{\star}}\right|^2=P,\forall k\in\mathcal{K}_t.
\end{align}

\begin{IEEEproof}
The proof of Lemma $1$ is omitted due to the space limitation.
\end{IEEEproof}

As we can see from Lemma 1, we obtain the solution for $v_k$ by considering the following two cases: 1) When the transmit power budget is sufficiently large, we choose $v_k$ to force the value of objective function to be zero; 2) when the transmit power budget is limited, $v_k$ is optimized by fully exploiting the transmit power, namely the equality of transmit power constraint needs to be satisfied.

\subsection{Optimization of the Diagonal Reflect and Transmit Beamforming Vectors $\bm{\theta}_r$, $\bm{\theta}_t$}
By fixing $\{v_k\}$ and $\bu$, the subproblem with respect to $\bm{\theta}_r$ and $\bm{\theta}_t$ is given by
\begin{equation}\label{thetapro}
\begin{split}
\mathcal{P}6\!:\;&\opmin_{\bm{\theta}_r,\bm{\theta}_t} ~ \zeta\\
&\text{subject to} ~ \bm{\theta}_r(m)=e^{j\theta_m^r},\forall m\in\mathcal{M},\\
&~~~~~~~~~~~~\bm{\theta}_t(m)=e^{j\theta_m^t},\forall m\in\mathcal{M},\\
&~~~~~~~~~~~~|\theta_m^t-\theta_m^r|=\frac{\pi}{2}~\text{or}~\frac{3\pi}{2}, \forall m\in\mathcal{M},\\
&~~~~~~~~~~~~\theta_m^r,\theta_m^t\in[0,2\pi], \forall m\in\mathcal{M}.
\end{split}
\end{equation}
It is clear that problem $\mathcal{P}6$ is challenging due to the coupled phase-shift constraints and the constant modulus constraints. In the following, we first reformulate problem $\mathcal{P}6$ to a more tractable form by introducing an auxiliary binary vector and exploiting the coupled binary phase-shift constraints. Then the constant modulus constraints can be exploited to solve this relaxed problem by updating the elements in $\bm{\theta}_r$ and the introduced binary vector separately. Finally, $\bm{\theta}_t$ is optimized by considering the coupled phase-shift constraints.

From the coupled phase-shift constraints, it can be inferred that $\theta_m^t = \theta_m^r\pm \frac{\pi}{2}$ or $\frac{3\pi}{2}$. Then, we have
\begin{align}
\bm{\theta}_t(m)=e^{j\theta_m^t}=\bm{q}(m)\sqrt{-1}e^{j\theta_m^r}=\bm{q}(m)\sqrt{-1}\bm{\theta}_r(m),
\end{align}
where $\bm{q}\in\Rdom^{M\times 1}$ is a binary vector with $\bm{q}(m)\in\{-1,1\}$. Thus the relationship between $\theta_m^t$ and $\theta_m^r$ depends on the value of $\bm{q}$, leading to a binary phase-shift constraint. Therefore, problem $\mathcal{P}6$ is reformulated to
\begin{equation}\label{thetapro0}
\begin{split}
\mathcal{P}7\!:\;&\opmin_{\bm{\theta}_r,\bm{q}} ~ \hat{\zeta}\\
&\text{subject to} ~ \bm{\theta}_r(m)=e^{j\theta_m^r},\forall m\in\mathcal{M},\\
&~~~~~~~~~~~~\bm{q}(m)\in\{-1,1\},\forall m\in\mathcal{M},
\end{split}
\end{equation}
where
\begin{align}
\hat{\zeta}&=\sum_{k\in\mathcal{K}_r}|\bu^H\bh_kv_k+\bm{\theta}_r^H\text{diag}(\bu^H\bG)\bg_kv_k-1|^2\nonumber\\
&+\sum_{k\in\mathcal{K}_t}|\bu^H\bh_kv_k+(\bm{\theta}_r\circ \sqrt{-1}\bm{q})^H\text{diag}(\bu^H\bG)\bg_kv_k-1|^2\nonumber\\
&+\sigma^{2}\bu^H\bu.
\end{align}

One can see that, by denoting $\bm{\theta}_t$ as the Hadamard product of $\bm{\theta}_r$, $\bm{q}$, and the unit imaginary vector, the variables of problem $\mathcal{P}7$ are no longer coupled in the constraints and the challenge is only due to the constant modulus constraints. To deal with this issue, we apply the first-order optimality condition with constant modulus constraints. Since the variables $\bm{\theta}_r$ and $\bm{q}$ are separated in the constraints, we can address problem $\mathcal{P}7$ in an alternating manner. Besides, due to the fact that the elements of $\bm{\theta}_r$ are also separated, $\bm{\theta}_r$ can be optimized by updating $\bm{\theta}_r(m)$ in the order of $m\in\mathcal{M}$, i.e., solving the following subproblem:
\begin{equation}\label{thetapro1}
\begin{split}
\mathcal{P}8\!:\;&\opmin_{\bm{\theta}_r(m)} ~ \sum_{k\in\mathcal{K}_r}|\bm{\theta}^{*}_r(m)\ba_{k}(m)+\bb_{k}(m)|^2\\
&~~~~~~~~~~~+\sum_{k\in\mathcal{K}_t}|(\bm{\theta}_r(m)\sqrt{-1}\bm{q}(m))^{*}\bc_{k}(m)+\bd_{k}(m)|^2\\
&\text{subject to} ~ \bm{\theta}_r(m)=e^{j\theta_m^r},
\end{split}
\end{equation}
where
\begin{align}
&\ba_k = \text{diag}(\bu^H\bG)\bg_kv_k,\forall k\in\mathcal{K}_r,\\
&\bb_{k}(m) = \bu^H\bh_kv_k-1+\sum_{j\neq m}\bm{\theta}^{*}_r(j)\ba_k(j),\nonumber\\
&\forall k\in\mathcal{K}_r, \forall m,j\in\mathcal{M},\\
&\bc_k = \text{diag}(\bu^H\bG)\bg_kv_k,\forall k\in\mathcal{K}_t,\\
&\bd_k(m) = \bu^H\bh_kv_k-1+\sum_{j\neq m}(\bm{\theta}_r(j)\sqrt{-1}\bm{q}(j))^{*}\bc_k(j),\nonumber\\
&\forall k\in\mathcal{K}_t, \forall m,j\in\mathcal{M}.
\end{align}
Note that $|\bm{\theta}_r(m)|^2=1$. By ignoring some constant terms, problem $\mathcal{P}8$ can be equivalently rewritten as follows:
\begin{equation}\label{thetapro2}
\begin{split}
\mathcal{P}9\!:\;&\opmin_{\bm{\theta}_r(m)} ~ \Re[\bm{\theta}^{*}_r(m)\eta]\\
&\text{subject to} ~ \bm{\theta}_r(m)=e^{j\theta_m^r},
\end{split}
\end{equation}
where $\eta=\sum_{k\in\mathcal{K}_r}\ba_k(m)\bb^{*}_{k}(m)+\sum_{k\in\mathcal{K}_t}\sqrt{-1}\bm{q}^{*}(m)\bc_k(m)\bd_k^{*}(m)$. Without the constant modulus constraint, the minimum objective function value of problem $\mathcal{P}9$ can be obtained by choosing $\bm{\theta}_r(m) = -\eta$. Therefore, the feasible solution for $\bm{\theta}_r(m)$ is given by
\begin{align}
\bm{\theta}_r(m) = \frac{-\eta}{|\eta|}.\label{optthetar}
\end{align}

Similarly, $\bm{q}(m)$ can be updated element-wisely, which leads to the following problem:
\begin{equation}\label{thetapro3}
\begin{split}
\mathcal{P}10\!:\;&\opmin_{\bm{q}(m)} ~ \sum_{k\in\mathcal{K}_t}|(\bm{\theta}_r(m)\sqrt{-1}\bm{q}(m))^{*}\bc_{k}(m)+\bd_{k}(m)|^2\\
&\text{subject to} ~ \bm{q}(m)\in\{-1,1\}.
\end{split}
\end{equation}
Due to the fact that $\bm{q}(m)$ is a real variable and $|\bm{q}(m)|^2=1$, the objective function can be equivalently rewritten as $\bm{q}(m)\Re[(\bm{\theta}_r(m)\sqrt{-1})^{*}\bc_k(m)\bd^{*}_k(m)]$. It can be inferred that the objective function value is minimized by choosing $\bm{q}(m)=-1$ if $\Re[\sum_{k\in\mathcal{K}_t}(\bm{\theta}_r(m)\sqrt{-1})^{*}\bc_k(m)\bd^{*}_k(m)]\geq 0$; otherwise, $\bm{q}(m)=1$. Hence, $\bm{q}(m)$ is updated by
\begin{align}
\bm{q}(m) = -\text{sign}\{\Re[\sum_{k\in\mathcal{K}_t}(\bm{\theta}_r(m)\sqrt{-1})^{*}\bc_k(m)\bd^{*}_k(m)]\}.\label{optq}
\end{align}

Finally, $\bm{\theta}_t$ can be optimized by considering the binary phase-shift constraints, i.e.,
\begin{align}
\bm{\theta}_t = \bm{\theta}_r\circ (\sqrt{-1}\bm{q}).\label{optthetat}
\end{align}

According to the above derivation, the AO-BPC algorithm for solving problem $\mathcal{P}1$ is summarized in Algorithm \ref{AOBPC}. One can see that, by exploiting the alternating optimization scheme and some conventional methods (e.g., the first-order optimality condition and the Lagrange multiplier method), we derive the closed-form solutions for $\bu$ and $\{v_k\}$ and then address problems $\mathcal{P}2$ and $\mathcal{P}3$ efficiently. Beside, since the coupled phase-shift constraints between $\bm{\theta}_r$ and $\bm{\theta}_t$ in the subproblem $\mathcal{P}6$ are non-convex and uncertain, we introduce the auxiliary binary vector $\bm{q}$ and rewrite these constraints in the form of Hadamard product and reformulate $\mathcal{P}6$ into a more tractable problem with respect to $\bm{\theta}_r$ and $\bm{q}$, where the variables are separated in the constraints. Finally, to tackle the binary constraints in the reformulated problem with respect to $\bm{\theta}_r$ and $\bm{q}$, inspiring by the binary nature of $\bm{q}$, we update these variables with their closed-form solutions in an alternating manner. Then the solution of $\bm{\theta}_t$ can be derived by considering the Hadamard-product constraint.

\begin{algorithm}[t]

\caption{\label{AOBPC}The AO-BPC Algorithm for Solving Problem $\mathcal{P}1$.}

\textbf{Initialize }{$\epsilon>0$, $\{v_k\}$, $\bu$, $\bm{\theta}_r$, and $\bm{\theta}_t$ such that they meet all the constraints;}{\small\par}

\textbf{Repeat}

\textbf{\,\,\,\,\,}{Update $\bu$ according to \eqref{optu};}

\textbf{\,\,\,\,\,}{Update $\{v_k\}$ according to Lemma $1$;}

\textbf{\,\,\,\,\,}{Update $\bm{\theta}_r$ and $\bm{\theta}_t$ according to \eqref{optthetar}, \eqref{optq}, and \eqref{optthetat};}

\textbf{Until }{the decrease of the objective is below $\epsilon$.}
\end{algorithm}

\subsection{The Analysis of Complexity and Convergence}

Considering \eqref{optu}, the complexity of updating $\bu$ is $\mathcal{O}(K(M^2+MN+N)+N^3)$. Similarly, the complexity of optimizing $\{v_k\}$ is $\mathcal{O}(K(M^2+MN+N))$. One can see that, the complexities of optimizing $\bm{\theta}_r$ and $\bm{q}$ are both $\mathcal{O}(KM)$. The complexity of updating $\bm{\theta}_t$ based on \eqref{optthetat} is $\mathcal{O}(M)$. Then the overall complexity of Algorithm \ref{AOBPC} is $\mathcal{O}(K(M^2+MN+N)+N^3+KM+M)$. The convergence of Algorithm \ref{AOBPC} is also analyzed, which is summarized in the following theorem.

\theorem It is guaranteed that any limit point of the sequence generated by the proposed AO-BPC algorithm in Algorithm \ref{AOBPC} converges to a stationary solution of problem $\mathcal{P}1$.

\begin{IEEEproof}
See Appendix A.
\end{IEEEproof}

\section{Comparison With the Conventional RIS Assisted AirComp Systems}
In this section, we first derive the computation MSE for the conventional RIS assisted AirComp systems. Then under some reasonable assumptions, the computation MSEs for the STAR-RIS assisted and conventional RIS assisted systems are simplified. Finally, we theoretically prove that the STAR-RIS can provide more performance gain than the conventional RIS with respect to the computation MSE.

By replacing the STAR-RIS with a conventional RIS, the received signal at the FC is given by
\begin{align}
\tilde{y}=\sum_{k\in\mathcal{K}_r}(\bh_k+\bG\bm{\Theta}\bg_k)v_ks_k+\sum_{k\in\mathcal{K}t}\bh_kv_ks_k+\bn,
\end{align}
where $\bm{\Theta}\in\Cdom^{M\times M}$ is the passive beamforming matrix at the RIS. Note that since the conventional RIS cannot perform signal transmission, there are only direct links between the FC and the WDs located in the transmission space. By exploiting the receive beamforming vector $\tilde{\bu}\in\Cdom^{N\times 1}$, the computed result is expressed as
\begin{align}
\tilde{s}=\tilde{\bu}^H\tilde{y}.
\end{align}
Then the computation MSE for the conventional RIS assisted AirComp systems is formulate as follows:
\begin{align}
\tilde{\zeta} &= \mathbb{E}(|\tilde{s}-s|^2)\nonumber\\
&=\sum_{k\in\mathcal{K}_r}|\tilde{\bu}^H(\bh_k+\bG\bm{\Theta}\bg_k)v_k-1|^2 + \sum_{k\in\mathcal{K}_t}|\tilde{\bu}^H\bh_kv_k-1|^2\nonumber\\
&+\sigma^2\tilde{\bu}^H\tilde{\bu}.\label{MSEconven}
\end{align}

By assuming that $\tilde{\bu}$ is also optimized by the first-order optimality condition, then $\tilde{\bu}$ can be updated by
\begin{align}
\tilde{\bu} = &\left(\sum_{k\in\mathcal{K}_r}|v_k|^2\tilde{\bh}_{k,r}\tilde{\bh}_{k,r}^H+\sum_{k\in\mathcal{K}_t}|v_k|^2\bh_k\bh_k^H+\sigma^2\bI\right)^{-1}\nonumber\\
&\times\left(\sum_{k\in\mathcal{K}_r}\tilde{\bh}_{k,r}v_k+\sum_{k\in\mathcal{K}_t}\bh_kv_k\right).\label{optu0}
\end{align}
where $\tilde{\bh}_{k,r}=\bh_k+\bG\bm{\Theta}\bg_k$. By substituting \eqref{optu0} into \eqref{MSEconven}, the computation MSE can be rewritten as
\begin{align}
&\tilde{\zeta}(v_k,\bm{\Theta}) = K-\left(\sum_{k\in\mathcal{K}_r}\tilde{\bh}_{k,r}v_k+\sum_{k\in\mathcal{K}_t}\bh_kv_k\right)^H\nonumber\\
&\times\left(\sum_{k\in\mathcal{K}_r}|v_k|^2\tilde{\bh}_{k,r}\tilde{\bh}_{k,r}^H+\sum_{k\in\mathcal{K}_t}|v_k|^2\bh_k\bh_k^H+\sigma^2\bI\right)^{-1}\nonumber\\
&\times \left(\sum_{k\in\mathcal{K}_r}\tilde{\bh}_{k,r}v_k+\sum_{k\in\mathcal{K}_t}\bh_kv_k\right). \label{MSEconven0}
\end{align}
Similarly, the computation MSE for the STAR-RIS assisted AirComp systems is given by
\begin{align}
&\zeta(v_k,\bm{\Theta}_r,\bm{\Theta}_t) = K-\left(\sum_{k\in\mathcal{K}_r}\bh_{k,r}v_k+\sum_{k\in\mathcal{K}_t}\bh_{k,t}v_k\right)^H\nonumber\\
&\times\left(\sum_{k\in\mathcal{K}_r}|v_k|^2\bh_{k,r}\bh_{k,r}^H+\sum_{k\in\mathcal{K}_t}|v_k|^2\bh_{k,t}\bh_{k,t}^H+\sigma^2\bI\right)^{-1}\nonumber\\ &\times\left(\sum_{k\in\mathcal{K}_r}\bh_{k,r}v_k+\sum_{k\in\mathcal{K}_t}\bh_{k,t}v_k\right).\label{MSESTAR}
\end{align}

For the ease of analysis, we consider the following assumption:
\assump We assume that the number of antennas at the FC and the number of elements at the STAR-RIS/RIS are sufficiently large such that the channel vectors of different WDs are orthogonal, i.e.,
\begin{align}
&\frac{1}{N}\bh_k^H\bh_{k^{'}} = \bm{0}, ~\frac{1}{N}\bh_{j,r}^H\bh_{j^{'},r} = \bm{0}, \nonumber\\
&\forall k,k^{'}\in\mathcal{K},k\neq k^{'},\forall j,j^{'}\in\mathcal{K}_r,j\neq j^{'},\label{ass0}\\
&\frac{1}{N}\bh_{k,t}^H\bh_{k^{'},t} = \bm{0}, ~\frac{1}{N}\bh_{j,r}^H\bh_{j^{'},t} = \bm{0}, \nonumber\\
&\forall k,k^{'}\in\mathcal{K}_t,k\neq k^{'},\forall j\in\mathcal{K}_r, j^{'}\in\mathcal{K}_t,\label{ass1}\\
&\frac{1}{N}\tilde{\bh}_{k,r}^H\tilde{\bh}_{k^{'},r} = \bm{0}, ~\frac{1}{N}\tilde{\bh}_{j,r}^H\bh_{j^{'}} = \bm{0}, \nonumber\\
&\forall k,k^{'}\in\mathcal{K}_r,k\neq k^{'},\forall j\in\mathcal{K}_r,j^{'}\in\mathcal{K}_t.\label{ass2}
\end{align}

With the above assumption, we can simplify the computation MSEs in \eqref{MSEconven0} and \eqref{MSESTAR} to a more tractable form, leading to the following lemma.
\lemma With Assumption $1$, \eqref{MSEconven0} and \eqref{MSESTAR} can be approximately written as follows:
\begin{align}
&\tilde{\zeta}(v_k,\bm{\Theta})\approx K-\frac{1}{\sigma^2}\Bigg{(}\sum_{k\in\mathcal{K}_r}\frac{\sigma^2|v_k|^2\tilde{\bh}_{k,r}^H\tilde{\bh}_{k,r}}{\sigma^2+|v_k|^2\tilde{\bh}_{k,r}^H\tilde{\bh}_{k,r}}\nonumber\\
&~~~~~~~~~~~~~+\sum_{k\in\mathcal{K}_t}\frac{\sigma^2|v_k|^2\bh_{k}^H\bh_{k}}{\sigma^2+|v_k|^2\bh_{k}^H\bh_{k}}\Bigg{)},\label{MSEapp0}\\
&\zeta(v_k,\bm{\Theta}_r,\bm{\Theta}_t)\approx K-\frac{1}{\sigma^2}\Bigg{(}\sum_{k\in\mathcal{K}_r}\frac{\sigma^2|v_k|^2\bh_{k,r}^H\bh_{k,r}}{\sigma^2+|v_k|^2\bh_{k,r}^H\bh_{k,r}}\nonumber\\
&~~~~~~~~~~~~~+\sum_{k\in\mathcal{K}_t}\frac{\sigma^2|v_k|^2\bh_{k,t}^H\bh_{k,t}}{\sigma^2+|v_k|^2\bh_{k,t}^H\bh_{k,t}}\Bigg{)},\label{MSEapp1}
\end{align}
respectively.

\begin{IEEEproof}
See Appendix B.
\end{IEEEproof}

According to Lemma $2$, we theoretically prove that the computation MSE of STAR-RIS assisted AirComp system is less than that of conventional RIS assisted AirComp system, which is summarized in the following theorem.
\theorem Let $\tilde{v}_k^{\star}$ ($v_k^{\star}$) and $\tilde{\bm{\Theta}}^{\star}$ ($\bm{\Theta}_r^{\star},\bm{\Theta}_t^{\star}$) denote the optimal transmit coefficient at the WDs and passive beamforming matrix for conventional RIS assisted (STAR-RIS assisted) AirComp systems. The computation MSE of the STAR-RIS assisted AirComp system is less than that of the conventional RIS assisted system, i.e., $\zeta(v_k^{\star},\bm{\Theta}_r^{\star},\bm{\Theta}_t^{\star})< \tilde{\zeta}(\tilde{v}_k^{\star},\tilde{\bm{\Theta}}^{\star})$.

\begin{IEEEproof}
See Appendix C.
\end{IEEEproof}

\section{Simulation Results}

In this section, we evaluate the computation MSE performance of the proposed AO-BPC algorithm by comparing with the following benchmark schemes:
\begin{itemize}
\item\; CRIS: In this benchmark scheme, the STAR-RIS is replaced with the conventional RIS. Hence, the signal from T-WDs will not be processed by the RIS, which means that there are only direct links between the FC and the T-WDs. Hence, by excluding the update of the diagonal transmit beamforming vector, we can also optimize the receive beamforming vector at the FC, the transmit coefficients at the WDs, and the diagonal reflect vector at the RIS according to the AO-BPC algorithm.

\item\; AO-RPC: In this benchmark scheme, the random phase-shift constraint between $\bm{\theta}_t$ and $\bm{\theta}_r$ is considered, i.e., $\bm{\theta}_t=\bm{\theta}_r\circ \hat{\bm{q}}$ with $\hat{\bm{q}}\in\Cdom^{M\times 1}$ and $|\hat{\bm{q}}(m)|=1$. In particular, we assume that $\hat{\bm{q}}$ is fixed in the beamforming design. Then we can update $\bm{\theta}_r$ and $\bm{\theta}_t$ by solving a reformulated subproblem in the form of problem $\mathcal{P}8$ and considering the random phase-shift constraint.

\item\; AO-WPC: In this benchmark scheme, the phase-shift constraint between $\bm{\theta}_t$ and $\bm{\theta}_r$ is removed. Therefore, $\bm{\theta}_t$ and $\bm{\theta}_r$ are optimized separately by exploiting the method applied to problem $\mathcal{P}9$.
\end{itemize}

\begin{figure}[htbp]
\centering
\includegraphics[width=2.5in]{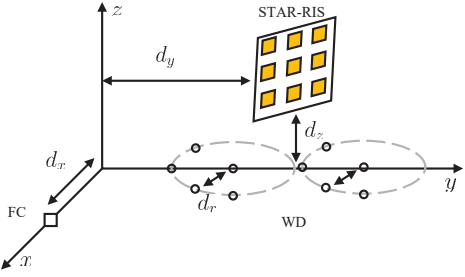}
\captionsetup{justification=raggedright}
\caption{The location setup in the simulations.}
\label{fig:location}
\end{figure}

In the following simulations, the tightly spaced antennas and the limited angular spread of scattering environment are assumed. Thus we consider a general spatially correlated Rician fading channel model for the WD--RIS, WD--FC, and RIS--FC links, which consists of both line-of-sight (LoS) and non-LoS components and is widely used in literature \cite{Mu2021}. In the LoS component, the distance-dependent path
loss is defined as $L=C_0\left(\frac{d}{D_0}\right)^{-\alpha}$ with
$C_0$, $d$, and $\alpha$ being the path loss at the reference distance $D_0=1$ meter
(m), the individual distance, and the path loss exponent. To be clear, we denote $\alpha_{\mathrm{WR}}$ ($\alpha_{\mathrm{rWF}}$/$\alpha_{\mathrm{tWF}}$ and $\alpha_{\mathrm{RF}}$) and $\beta_{\mathrm{WR}}$ ($\beta_{\mathrm{rWF}}$/$\beta_{\mathrm{tWF}}$ and $\beta_{\mathrm{SF}}$) as the path loss exponent and the Rician factor for the
WD--RIS (WD--FC and RIS-FC) links. In the simulation setup, a
three-dimensional system of Fig.~\ref{fig:location} is considered, where
the FC and the STAR-RIS are on the $x$-axis and $y$-$z$ plane,
respectively. The FC is assumed to equipped with a ULA, while an uniform rectangular array (URA) with $M=M_yM_z$ reflecting elements are deployed at the STAR-RIS with $M_y$ and $M_z$ denoting the number of elements along the $y$-axis and $z$-axis, respectively. The reference antenna/element at the FC/STAR-RIS are located at $(d_x=2\,\mathrm{m},0,0)$ and $(0,d_y=50\,\mathrm{m},d_z=3\,\mathrm{m})$. Besides, the R-WDs/T-WDs uniformly locate on a circle with radius
$d_r=5\,\mathrm{m}$ whose center is at $(0,d_y-d_r=45\,\mathrm{m}, 0)$/$(0,d_y+d_r=55\,\mathrm{m}, 0)$. We define the signal-to-noise-ratio (SNR)
as $\text{SNR}\,=\frac{P}{\sigma^2}$. Furthermore, the simulation parameter settings in Table \RNum{2} are considered unless otherwise specified.
\begin{table}[htbp]
\centering
\caption{The settings of the simulation parameters.}
\begin{tabular}{|c|c|c|c|}
\hline
Parameter& Value & Parameter& Value\\
\hline
$M$ & $64$ & $\alpha_{\mathrm{tWF}}$ & $4$\\
\hline
$M_y$ & $4$ & $\alpha_{\mathrm{RF}}$ & $3$\\
\hline
$M_z$ & $M/M_y$ & $d_x$ & $2$~m\\
\hline
$N$ & $64$ & $d_y$ & $50$~m\\
\hline
$K$ & $64$ & $d_z$ & $3$~m\\
\hline
$K_r$ & $K/2$ & $d_r$ & $3$~m\\
\hline
$K_t$ & $K/2$ & $\beta_{WS}$ & $3\,\text{dB}$\\
\hline
$\sigma^2$ & $-80\,\text{dBm}$ & $\beta_{SF}$ & $3\,\text{dB}$\\
\hline
$C_0$ & $-30\,\text{dB}$ & $\beta_{rWF}$ & $-3\,\text{dB}$\\
\hline
$D_0$ & $1$~m & $\beta_{tWF}$ & $-3\,\text{dB}$\\
\hline
$\alpha_{\mathrm{WR}}$ & $2.2$ & $\epsilon$ & $10^{-4}$\\
\hline
$\alpha_{\mathrm{rWF}}$ & $3.8$ &  & \\
\hline
\end{tabular}
\end{table}
\vspace{-.3cm}
\subsection{Convergence}
\begin{figure}[htbp]
\centering
\includegraphics[width=2.5in]{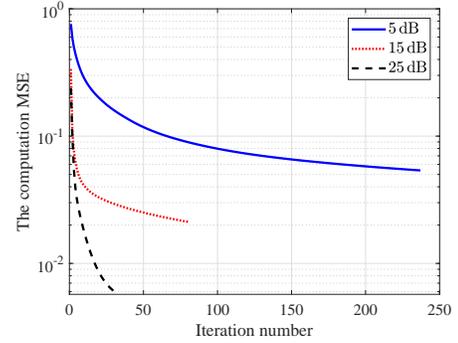}
\captionsetup{justification=raggedright}
\caption{Convergence behaviour of the AO-BPC algorithm with $\text{SNR}\,=5\,\text{dB}$, $15\,\text{dB}$, and $25\,\text{dB}$.}
\label{fig:fig2}
\end{figure}
Fig. \ref{fig:fig2} shows the convergence behaviour of the AO-BPC algorithm with $\text{SNR}\,=5\,\text{dB}$, $15\,\text{dB}$, and $25\,\text{dB}$. In different SNR regions, the proposed algorithm converges to a stationary point after a number of iterations. The stationary point decreases when SNR increases and tends to be zero. Furthermore, in the moderate and high SNR region, the AO-BPC algorithm terminates within only a few iterations. However, one can see that the convergence speed slows down as the SNR becomes extremely low, which takes over $200$ iterations to converge with $\text{SNR}\,=5\,\text{dB}$.

\subsection{Impact of the Number of WDs}
\begin{figure}[htbp]
\centering
\includegraphics[width=2.5in]{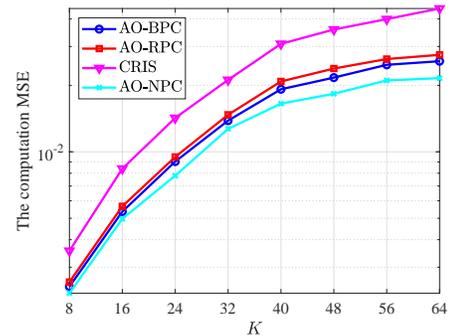}
\captionsetup{justification=raggedright}
\caption{The computation MSE performance versus $K$ with $\text{SNR}\,=15\,\text{dB}$.}
\label{fig:fig3}
\end{figure}
Fig. \ref{fig:fig3} compares the computation MSE performance of the considered schemes versus $K$ with $\text{SNR}\,=15\,\text{dB}$. One can see that the computation MSEs of all the schemes increase with the number of WDs due to the fact that it is more challenging to design a single passive reflect/transimit matrix at the STAR-RIS/RIS and a single receive beamforming vector at the FC to aggregate the collected data from more WDs. Besides, the AO-BPC algorithm outperforms the CRIS algorithm with an increasing gap. This coincides with our discussion in Section \RNum{4} that the STAR-RIS assisted AirComp system shows less computation MSE than the conventional RIS assisted system. The computation MSE achieved by the AO-BPC algorithm is less than that achieved by the AO-RPC algorithm with a minor gap, since these two algorithms both consider the coupled phase-shift constraints. The phase-shift constraints in the AO-BPC algorithm are binary, which can be decoupled by introducing an auxiliary binary vector. This vector can be optimized by exploiting its binary nature to improve the system performance. However, there lack efficient methods to address the random phase-shift constraints in the AO-RPC algorithm, which leads to the minor gap in Fig. \ref{fig:fig3}. Also, the random coupled phase-shift constraints may lead to high complexity and take more iterations to converge. The proposed algorithm approaches the AO-NPC algorithm, which can be viewed as a lower bound since the passive reflect and transmit beamforming matrices at the STAR-RIS are optimized by ignoring the coupled phase-shift constraints, which provides more spatial degrees. This further verifies the effectiveness of our proposed algorithm.

\subsection{Impact of the Number of R-WDs}
\begin{figure}[htbp]
\centering
\includegraphics[width=2.5in]{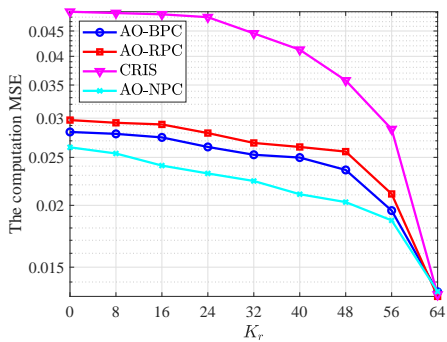}
\captionsetup{justification=raggedright}
\caption{The computation MSE performance versus $K_r$ with $K=64$ and $\text{SNR}\,=15\,\text{dB}$.}
\label{fig:fig4}
\end{figure}
We compare the computation MSE performance versus $K_r$ of the considered schemes with $K=64$ and $\text{SNR}\,=15\,\text{dB}$. As we can see, as $K_r$ increases, the computation MSEs of all the schemes decrease since more WDs become R-WDs with less pathloss according to Fig. \ref{fig:location}. Also, these R-WDs can benefit from the passive reflection matrix at the STAR-RIS/RIS. In particular, the effect of the passive reflection matrix is much more obvious in the CRIS algorithm since the conventional RIS works only when the WDs is located at the reflection space. When $K_r=0$, i.e., all the WDs are T-WDs, the STAR-RIS assisted AirComp system shows much less computation MSE than the conventional RIS assisted system. This is due to the fact that the signals from T-WDs are only transmitted via direct links between the T-WDs and the FC in the conventional RIS assisted systems since the RIS can only perform signal reflection, while in the STAR-RIS assisted system the signals can be transmitted by the STAR-RIS. Moreover, the computation MSEs achieved by the considered schemes coincide when $K_r=64$. In this case, all the WDs become R-WDs and the STAR-RIS does not perform signal transmission, i.e., the STAR-RIS assisted AirComp systems reduce to the conventional RIS assisted ones.

\subsection{Impact of the Number of Passive Reflecting/Transmiting Elements at the STAR-RIS/RIS}
\begin{figure}[htbp]
\centering
\includegraphics[width=2.5in]{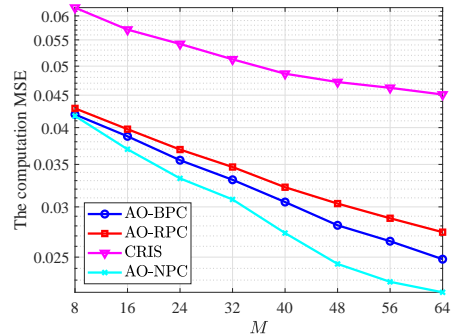}
\captionsetup{justification=raggedright}
\caption{The computation MSE performance versus $M$ with $\text{SNR}\,=15\,\text{dB}$.}
\label{fig:fig5}
\end{figure}
Fig. \ref{fig:fig5} shows the computation MSE performance of the considered schemes versus the number of passive reflecting/transmiting elements at the STAR-RIS/RIS with $\text{SNR}\,=15\,\text{dB}$. As $M$ increases, the STAR-RIS/RIS can provide more channel gain and then the computation MSEs decrease. The AO-BPC algorithm outperforms the AO-RPC and CRIS algorithms with increasing gaps due to the facts that the former can not only offer more spatial degrees for the T-WDs (i.e., signal transmission) but also better deal with the coupled phase-shift constraints between the passive reflect and transmit beamforming matrices at the STAR-RIS when $M$ becomes larger. Also, the computation MSE of our proposed algorithm is lower bounded by that achieved by the AO-NPC algorithm. But the gap is quire small especially when $M$ is less than $32$.

\subsection{Impact of the Number of Receive Antennas at the FC}
\begin{figure}[htbp]
\centering
\includegraphics[width=2.5in]{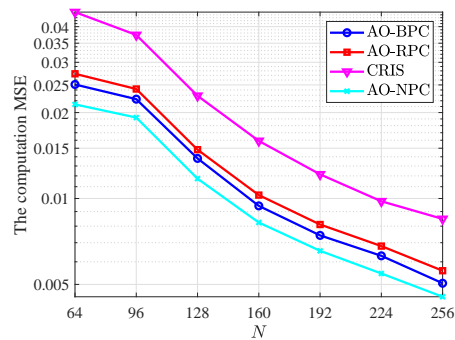}
\captionsetup{justification=raggedright}
\caption{The computation MSE performance versus $N$ with $\text{SNR}\,=15\,\text{dB}$.}
\label{fig:fig6}
\end{figure}
Fig. \ref{fig:fig6} compares the computation MSE performance versus the number of receive antennas at the FC with $\text{SNR}\,=15\,\text{dB}$. Since more receive antennas provide more spatial degrees, the computation MSEs of all the curves decrease as $N$ increases. Specially, the computation MSE decreases slowly when $N$ increases from $64$ to $96$, while the speed becomes faster when $N>96$. It is due to the fact that the channel gain of direct links between the WDs and the FC tends to be dominant with sufficiently large number of receive antennas. The proposed AO-BPC algorithm still outperforms the AO-RPC and CRIS algorithms. Specifically, the gap between the AO-BPC and CRIS algorithms is much larger than that between the AO-BPC and AO-RPC algorithms, which verifies the outstanding advantages of STAR-RIS in AirComp systems.

\subsection{Impact of the SNR}
\begin{figure}[htbp]
\centering
\includegraphics[width=2.5in]{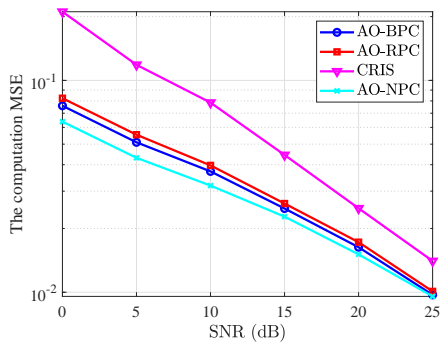}
\captionsetup{justification=raggedright}
\caption{The computation MSE performance versus SNR.}
\label{fig:fig7}
\end{figure}
Fig. \ref{fig:fig7} presents the computation MSE of the considered schemes versus SNR. As SNR increases, the computation MSE performance improves significantly due to the channel gain. Besides, it can be observed that the gap between the AO-BPC and AO-NPC algorithms turn to be smaller when SNR changes from $0\,\text{dB}$ to $25\,\text{dB}$. It means that though the binary phase-shift constraints are considered in the AO-BPC algorithm, the computation MSE achieved by our proposed algorithm is quite close to that achieved by the AO-NPC algorithm in the high-SNR region, which is viewed as the performance lower bound. Such observation further verifies the efficiency of the AO-BPC algorithm.

\section{Conclusion}
In this work, the joint beamforming design for the STAR-RIS assisted AirComp systems was investigated. An AO-BPC algorithm was proposed to jointly optimize the transmit coefficients at the R-WDs/T-WDs, the passive reflect/transmit beamforming matrices at the STAR-RIS, and the receive beamforming vector at the FC. In particular, we optimized the transmit coefficients and the receive beamforming vector by applying the Lagrange duality method and the first-order optimality condition. By introducing an auxiliary variable and exploiting the binary phase-shift constraints, we derived the closed-form solutions for the passive reflect and transmit beamforming matrices. The analysis of complexity and convergence for the proposed AO-BPC algorithm has been also presented. Moreover, under some reasonable assumptions, the expressions of the computation MSEs for the STAR-RIS assisted and conventional RIS assisted AirComp systems are simplified. Then it is proven that the STAR-RIS assisted AirComp system can achieve less computation MSE than the conventional RIS assisted system. Our simulation results showed that our proposed AO-BPC algorithm outperforms the AO-RPC and CRIS benchmark schemes. Besides, the computation MSE of the AO-BPC algorithm is close to the performance lower bound achieved by optimizing the passive reflect and transmit beamforming matrices without coupled phase-shift constraints. Our proposed AO-BPC algorithm can serve as an excellent candidate for STAR-RIS assisted AirComp system.

In the following, we discuss some issues that are not addressed yet in this work to motivate future research.
\begin{itemize}
\item\; As an initial attempt, we only considered the energy splitting protocol in this work. In general, STAR-RIS can operate in different protocols, e.g., mode switching and time switching. The performance and complexity of beamforming design may be significantly affected by the protocols. Therefore, it is important to investigate the STAR-RIS assisted AirComp systems under different protocols in the future.
\item\; In this work, the CSI is assume to be perfectly known at the WDs, the FC, and the STAR-RIS. In practise, since the number of elements at the STAR-RIS is large, it is challenging to estimate channel information and the error may be large. Hence, the robust beamforming design for STAR-RIS assisted AirComp systems is worthy of further investigation.
\item\; Considering the hardware limitation, the STAR-RIS may only perform discrete phase shift in practice. Therefore, the beamforming design for STAR-RIS assisted AirComp systems with discrete phase shift is also an important topic.
\end{itemize}

\appendices{
\section{Proof of Theorem 1}
Since the alternating optimization method is exploited to solve problem $\mathcal{P}1$, we need to prove that the objective function value is non-increasing in each iteration. Obviously, considering the fact that the receive beamforming vector $\bu$ at the FC and the transmit beamforming coefficients $\{v_k\}$ are updated based on the first-order optimality condition and Lagrange duality method, respectively, we have
\begin{align}
&\zeta(\bu_{n+1},\{v_{k,n}\},\bm{\theta}_{r,n},\bm{\theta}_{t,n})\leq \zeta(\bu_{n},\{v_{k,n}\},\bm{\theta}_{r,n},\bm{\theta}_{t,n}),\label{conpro0}\\
&\zeta(\bu_{n},\{v_{k,n+1}\},\bm{\theta}_{r,n},\bm{\theta}_{t,n})\leq \zeta(\bu_{n},\{v_{k,n}\},\bm{\theta}_{r,n},\bm{\theta}_{t,n}),\label{conpro1}
\end{align}
where $n$ denotes the iteration number. Therefore, the optimization of $\bu$ and $\{v_k\}$ leads to the non-increasing objective function value.

Next, let us focus on problem $\mathcal{P}7$. By considering the constant modulus constraints, we update the diagonal reflect vector $\bm{\theta}_r$ at the STAR-RIS and the auxiliary variable $\bm{q}$ element-wisely according to \eqref{optthetar} and \eqref{optq}, which are the optimal and feasible solutions for problems $\mathcal{P}9$ and $\mathcal{P}10$, respectively. Furthermore, the objective functions of $\mathcal{P}7$, $\mathcal{P}9$, and $\mathcal{P}10$ are all equivalent. Then we have
\begin{align}
&\hat{\zeta}(\bu_{n},\{v_{k,n}\},\bm{\theta}_{r,n+1},\bm{q}_{n})\leq \hat{\zeta}(\bu_{n},\{v_{k,n}\},\bm{\theta}_{r,n},\bm{q}_{n}),\label{conpro2}\\
&\hat{\zeta}(\bu_{n},\{v_{k,n}\},\bm{\theta}_{r,n},\bm{q}_{n+1})\leq \hat{\zeta}(\bu_{n},\{v_{k,n}\},\bm{\theta}_{r,n},\bm{q}_{n}).\label{conpro3}
\end{align}
Note that the diagonal transmit vector $\bm{\theta}_t$ is optimized by \eqref{optthetat}, which satisfies the coupled phase-shift constraints. Thus we have
\begin{align}
\hat{\zeta}(\bu_{n},\{v_{k,n}\},\bm{\theta}_{r,n},\bm{q}_{n})=\zeta(\bu_{n},\{v_{k,n}\},\bm{\theta}_{r,n},\bm{\theta}_{t,n}).\label{conpro4}
\end{align}
Combining \eqref{conpro0}-\eqref{conpro4}, the objective function value is proven to be non-increasing in each iteration. Besides, since the objective function is defined as the computation MSE, its value is lower-bounded by zero. Hence, the convergence of Algorithm \ref{AOBPC} follows immediately.

\section{Proof of Lemma 2}
We first focus on the simplification of \eqref{MSEconven0}. As we can see, the matrix inverse term is the most challenging to deal with. By exploiting its special structure, we can apply the Sherman-Morrison formula \cite{Mtxbook}. The matrix inverse term can be rewritten as
\begin{equation}
\bD_n =\left\{
\begin{split}
&\sigma^2\bI,&~\text{if}~n=0,\\
&\bD_{n-1}+|v_n|^2\tilde{\bh}_{n,r}\tilde{\bh}_{n,r}^H,&~\text{else if}~1<n\leq K_r,\\
&\bD_{n-1}+|v_n|^2\bh_{n}\bh_{n}^H,&~\text{otherwise}.
\end{split}
\right.
\end{equation}
Then we have the following recursive formula for calculate the inverse of $\bD_n$
\begin{equation}
\bD_n^{-1} = \left\{
\begin{split}
&\bD_{n-1}^{-1}-\frac{\bD_{n-1}^{-1}|v_n|^2\tilde{\bh}_{n,r}\tilde{\bh}_{n,r}^H\bD_{n-1}^{-1}}{1+|v_n|^2\tilde{\bh}_{n,r}^H\bD_{n-1}^{-1}\tilde{\bh}_{n,r}},\\
&~\text{if}~1\leq n\leq K_r\\
&\bD_{n-1}^{-1}-\frac{\bD_{n-1}^{-1}|v_n|^2\bh_n\bh_n^H\bD_{n-1}^{-1}}{1+|v_n|^2\bh_n^H\bD_{n-1}^{-1}\bh_n},\\
&~\text{otherwise}.
\end{split}
\right.
\end{equation}
By assuming that $K_r\geq 2$ and letting $n=1$, we can obtain the inverse of $\bD_1$
\begin{align}
\bD_1^{-1} = \frac{1}{\sigma^2}\left(\bI-\frac{|v_1|^2\tilde{\bh}_{1,r}\tilde{\bh}_{1,r}^H}{\sigma^2+|v_1|^2\tilde{\bh}_{1,r}^H\tilde{\bh}_{1,r}}\right).
\end{align}
Then we calculate the inverse of $\bD_2$
\begin{align}
&\bD_2^{-1} = \bD_1^{-1} - \frac{\bD_1^{-1}|v_2|^2\tilde{\bh}_{2,r}\tilde{\bh}_{2,r}^H\bD_1^{-1}}{1+|v_2|^2\tilde{\bh}_{2,r}^H\bD_1^{-1}\tilde{\bh}_{2,r}/\sigma^2}\nonumber\\
&\approx \frac{1}{\sigma^2}\left(\bI-\frac{|v_1|^2\tilde{\bh}_{1,r}\tilde{\bh}_{1,r}^H}{\sigma^2+|v_1|^2\tilde{\bh}_{1,r}^H\tilde{\bh}_{1,r}}-\frac{|v_2|^2\tilde{\bh}_{2,r}\tilde{\bh}_{2,r}^H}{\sigma^2+|v_2|^2\tilde{\bh}_{2,r}^H\tilde{\bh}_{2,r}}\right),
\end{align}
where the approximation is due to \eqref{ass2}. Therefore, we can derive the inverse of $\bD_K$ as follows
\begin{align}
\bD_K^{-1} \approx& \frac{1}{\sigma^2}\Bigg{(}\bI-\sum_{k\in\mathcal{K}_r}\frac{|v_k|^2\tilde{\bh}_k\tilde{\bh}_k^H}{\sigma^2+|v_k|^2\tilde{\bh}_{k,r}^H\tilde{\bh}_{k,r}}\nonumber\\
&-\sum_{k\in\mathcal{K}_t}\frac{|v_k|^2\bh_k\bh_k^H}{\sigma^2+|v_k|^2\bh_k^H\bh_k}\Bigg{)}.\label{invserterm}
\end{align}
By substituting \eqref{invserterm} into \eqref{MSEconven0}, we have
\begin{align}
&\tilde{\zeta}(v_k,\bm{\Theta}) \approx K-\left(\sum_{k\in\mathcal{K}_r}\tilde{\bh}_{k,r}v_k+\sum_{k\in\mathcal{K}_t}\bh_kv_k\right)^H\nonumber\\
&\times\bD_K^{-1}\left(\sum_{k\in\mathcal{K}_r}\tilde{\bh}_{k,r}v_k+\sum_{k\in\mathcal{K}_t}\bh_kv_k\right)\nonumber\\
&\approx K-\frac{1}{\sigma^2}\Bigg{(}\sum_{k\in\mathcal{K}_r}|v_k|^2\tilde{\bh}_{k,r}^H\tilde{\bh}_{k,r}+\sum_{k\in\mathcal{K}_t}|v_k|^2\bh_k^H\bh_k\nonumber\\
&-\sum_{k\in\mathcal{K}_r}\frac{|v_k|^4|\tilde{\bh}_{k,r}^H\tilde{\bh}_{k,r}|^2}{\sigma^2+|v_k|^2\tilde{\bh}_{k,r}^H\tilde{\bh}_{k,r}}-\sum_{k\in\mathcal{K}_t}\frac{|v_k|^4|\bh_k^H\bh_k|^2}{\sigma^2+|v_k|^2\bh_k^H\bh_k}\Bigg{)}\nonumber\\
&=K-\frac{1}{\sigma^2}\Bigg{(}\sum_{k\in\mathcal{K}_r}\frac{\sigma^2|v_k|^2\tilde{\bh}_{k,r}^H\tilde{\bh}_{k,r}}{\sigma^2+|v_k|^2\tilde{\bh}_{k,r}^H\tilde{\bh}_{k,r}}\nonumber\\
&+\sum_{k\in\mathcal{K}_t}\frac{\sigma^2|v_k|^2\bh_k^H\bh_k}{\sigma^2+|v_k|^2\bh_k^H\bh_k}\Bigg{)},
\end{align}
where the second approximation is due to \eqref{ass1} and \eqref{ass2}. \eqref{MSEapp1} can be obtained similarly and the results follows immediately.

\section{Proof of Theorem 2}
First, we assume that the STAR-RIS assisted AirComp systems also adopt the optimal transmit coefficients and passive reflect beamforming matrix of the conventional RIS assisted system. Then we have
\begin{align}
&\zeta(\tilde{v}_k^{\star},\tilde{\bm{\Theta}}^{\star},\bm{\Theta}_t)-\tilde{\zeta}(\tilde{v}_k^{\star},\tilde{\bm{\Theta}}^{\star})\nonumber\\
&=\frac{1}{\sigma^2}\sum_{k\in\mathcal{K}_t}\left(\frac{\sigma^2|v_k|^2\bh_k^H\bh_k}{\sigma^2+|v_k|^2\bh_k^H\bh_k}-\frac{\sigma^2|v_k|^2\bh_{k,t}^H\bh_{k,t}}{\sigma^2+|v_k|^2\bh_{k,t}^H\bh_{k,t}}\right)\nonumber\\
&=\frac{1}{\sigma^2}\sum_{k\in\mathcal{K}_t}\left(\frac{\sigma^2e_k}{\sigma^2+e_k}-\frac{\sigma^2e_k+f_k}{\sigma^2+e_k+f_k}\right),
\end{align}
where $e_k=|v_k|^2\bh_k^H\bh_k$ and $f_k=\sigma^2(2\Re(\bh_k^H\bG\bm{\Theta}_t\bg_k)+\bg_k^H\bm{\Theta}_t^H\bG^H\bG\bm{\Theta}_t\bg_k)$. Clearly, we can choose $\bm{\Theta}_t$=$\bar{\bm{\Theta}}_t$ such that the term $\Re(\bh_k^H\bG\bar{\bm{\Theta}}_t\bg_k)$ is positive. Then $f_k$ is positive. Since we have $\frac{a}{b}<\frac{a+c}{b+c}$ for any positive numbers $a,b,c$, the following inequality holds
\begin{align}
\frac{\sigma^2e_k}{\sigma^2+e_k}<\frac{\sigma^2e_k+f_k}{\sigma^2+e_k+f_k},
\end{align}
which leads to $\zeta(\tilde{v}_k^{\star},\tilde{\bm{\Theta}}_r^{\star},\bar{\bm{\Theta}}_t)<\tilde{\zeta}(\tilde{v}_k^{\star},\tilde{\bm{\Theta}}^{\star})$. Note that $\zeta(v_k^{\star},\bm{\Theta}_r^{\star},\bm{\Theta}_t^{\star})\leq \zeta(\tilde{v}_k^{\star},\tilde{\bm{\Theta}}^{\star},\bar{\bm{\Theta}}_t)$ always holds, therefore the results follow immediately.
}

\end{document}